# Photoluminescence switching effect in a two-dimensional atomic crystal

Zheng Sun[1,2*‡], Ke Xu[3,4,5‡], Chang Liu[1,6‡] Jonathan Beaumariage[2], Jierui Liang[2], Susan K Fullerton-Shirey[3], Zhe-Yu Shi[1], Jian Wu[1,7,8], and David Snoke[2]

[1]State Key Laboratory of Precision Spectroscopy, East China Normal University, Shanghai, 200241, China

[2]Department of Physics and Astronomy, University of Pittsburgh, Pittsburgh, PA 15260, USA

[3]Department of Chemical and Petroleum Engineering, University of Pittsburgh, Pittsburgh, PA 15260, USA

[4]School of Physics and Astronomy, Rochester Institute of Technology, 14623, USA

[5]Microsystems Engineering, Rochester Institute of Technology, 14623, USA

[6]Institute for Advanced Study, Tsinghua University, Beijing 100084, China

[7]Collaborative Innovation Center of Extreme Optics, Shanxi University, Taiyuan, Shanxi 030006, China

[8]CAS Center for Excellence in Ultra-intense Laser Science, Shanghai 201800, China

**ABSTRACT:**

Two-dimensional materials are an emerging class of new materials with a wide range of electrical and optical properties and potential applications. Single-layer structures of semiconducting transition metal dichalcogenides are gaining increasing attention for use in field-effect transistors. Here, we report a photoluminescence switching effect based on single-layer $WSe_2$ transistors. Dual gates are used to tune the photoluminescence intensity. In particular, a side-gate is utilized to control the location of ions within a solid polymer electrolyte to form an electric double layer at the interface of electrolyte and $WSe_2$ and induce a vertical electric field. Additionally, a back-gate is used to apply a 2nd vertical electric field. An on-off ratio of the light emission up to 90 was observed under constant pump light intensity. In addition, a blue shift of the photoluminescence line up to 36 meV was observed. We attribute this blue shift to the decrease of exciton binding energy due to the change of nonlinear in-plane dielectric constant, and use it to determine the 3rd order off-diagonal susceptibility $\chi^{(3)} = 3.50 \times 10^{-19}$ m$^2$/V$^2$.

* Email: zsun@lps.ecnu.edu.cn.

‡ Contributed equally to this work.

Low-dimensional semiconducting transition metal dichalcogenides (TMDs), with $MX_2$ stoichiometry, where M is a transition metal element from group VI (M = Mo, W) and X is a chalcogen (X = S, Se), have emerged as promising materials, offering complementary characteristics to graphene for electronics, photonics and optoelectronics applications because of their unusual electrical and optical properties.[1–6] For example, these monolayers can be readily assembled together like “Lego blocks,” without large lattice mismatch effects, due to the inter-plane van der Waals forces, which offer a convenient and flexible approach to design various devices[7]. In addition, TMDs can be integrated into photonic devices such as modulators, detectors, optical microcavities, etc.[8–12] Furthermore, the high mobility and the tunable bandgap of these materials enables them to be versatile components for electrical and optical circuits.[13–16]

Here we report a photoluminescence switching effect based on monolayer $WSe_2$ transistors and an observation of a large blue shift of the photoluminescence line up to 36 meV. The photoluminescence (PL) intensity of the $WSe_2$ element was tuned by applying vertical electric fields in a dual-gate geometry. A back-gate is used to apply an electric field from the bottom of the sample; a side-gate geometry and the electric double layer (EDL) gating method were used to maximize electric field at the top interface of $WSe_2$. When a side-gate voltage is applied, ions within a solid polymer electrolyte accumulate at the interface of the $WSe_2$ and induced countercharges. This design has the benefit that the interface electric field occurs across an extremely small distance, leading to a large electric field (> $10^7$ V/cm) and high capacitance density (1~10 μF/cm$^2$).[17–19] Another advantage of using EDL gating with side-gate geometry is that it does not block observation from above the channel, as a conventional top gate does, and therefore allows direct PL measurements from the top, as well as convenient integration into optoelectronic devices such as photodetectors or photoemitters[20].

## RESULTS AND DISCUSSION

We fabricated monolayer $WSe_2$ field effect transistors (FETs) with two different types of the source materials, one which was exfoliated and the other grown by chemical vapor deposition (CVD). Figure 1(a) shows typical PL spectra carried out before and after capping with the electrolyte polyethylene oxide (PEO): $CsClO_4$. The capping led to a 17-fold increase of the PL intensity and a narrowing of the full-width at half maximum (FWHM) of 6 meV. The increase of the PL intensity may be explained by the motion of the ions to smooth the extrinsic disorder, for example, helping to neutralize charged impurities in the $WSe_2$.[21] The electrolyte coverage may also help to prevent the $WSe_2$ from oxidizing in the air.

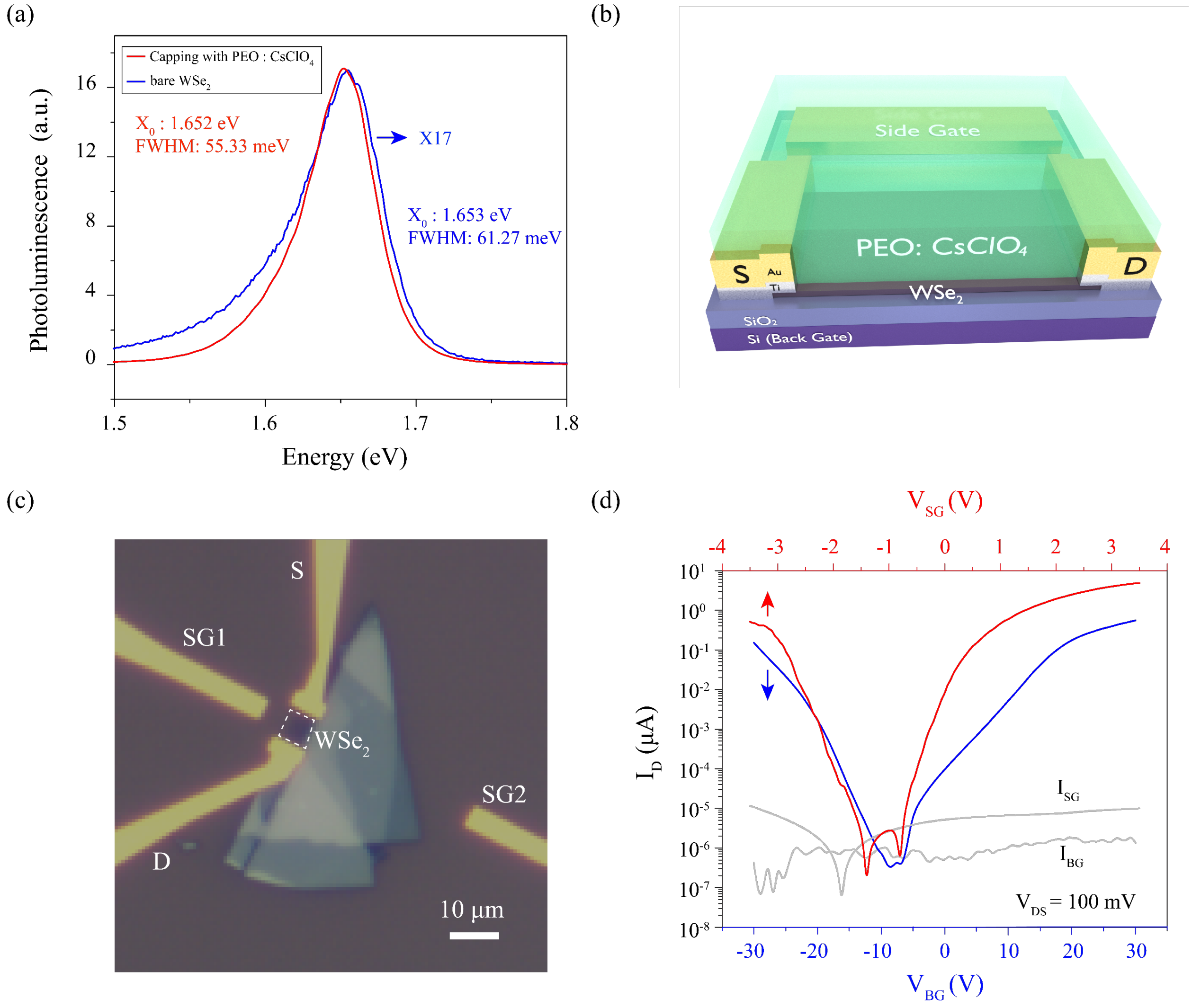

Fig. 1. (a) Typical photoluminescence spectrum of the exfoliated monolayer $WSe_2$ with (red) and without (blue) capping PEO: $CsClO_4$. A factor of 17 enhancement was obtained after coating along with decreasing the linewidth from 61.27 to 55.33 meV. (b) Schematic of the $WSe_2$ based transistors. S is the source, D is the drain, SG1 is side gate 1, and SG2 is side gate 2 (which was not used for these studies). The source-drain channel length was about 10 μm. (c) Optical microscope image of $WSe_2$-based transistor with two side-gates. (SG2 was not used for these measurements.) (d) Transfer characteristics of the FET before (blue) and after (red) depositing the PEO: $CsClO_4$ electrolyte. The back-gate scanning rate is 4 V/s and side gate scanning rate is 10 mV/s. $V_{DS}$ was kept at 100 mV. $I_{SG}$ and $I_{BG}$ are the leakage currents of the side-gate and the back-gate, respectively.

For the first type of sample, an isolated monolayer $WSe_2$, was mechanically exfoliated, with a typical size of 5×15 $\mu m^2$, from which a transistor was made. Figure 1(b) illustrates the transistor design, and Figure 1(c) shows an optical microscope image. A similar device was made using a CVD-grown $WSe_2$ layer; the optical microscope image is shown in the supplementary file Figure S1. The results of three-terminal transfer measurements before and after depositing the electrolyte on the exfoliated sample are shown in Figure 1(d). A typical ambipolar transport behavior is observed in the transistor based on exfoliated $WSe_2$, suggesting that electrons and the holes can both be accumulated depending on the polarity of applied gate voltage. The CVD-grown $WSe_2$ was p-doped, with only the hole-conduction branch observable in back-gate transfer measurements prior to the deposition of electrolyte (see Figure S2). As seen in Figure 1(d), using the electrolyte (red line) gave a 10-fold increase in the "On" current (n-branch), with only 1/10 of the applied voltage, thanks to the formation of the EDL and the resulting larger interface electric field.

After the initial optical and electrical characterization, we then performed gate-voltage-dependent PL measurements on the exfoliation sample. To investigate the gate-voltage dependence of the PL, we first applied -2 V on the side-gate (SG1) and monitored the change of the $WSe_2$ channel conductance with a small (100 mV) drain terminal voltage, while the source terminal was grounded. In response to the gate electric field, the cations and the anions within the electrolyte redistribute.[22,23] The channel conductance increased during this time as a result of the increase of the charge carriers (i.e., holes) in the $WSe_2$ layer. After the ions reach equilibrium, we then performed the PL measurements and monitored the PL intensity variations as we tuned the back-gate voltage.  All the measurements were performed in an ambient atmosphere at room temperature.

Figure 2 shows an unambiguous switching behavior, in which the PL intensity is turned off and turned on by modifying the back-gate voltages. PL intensity on/off ratios of 90 and 37 were observed in the exfoliated and CVD-grown samples, respectively, under the same conditions (the data for the CVD device is shown in the supplementary file Figure S3). For each voltage, we hold the voltage for 1 minute before PL measurements to allow the device reach equilibrium. The difference of PL intensity between initial, third, and fifth PL measurements with zero backgate voltage can potentially be attributed to the lagging effect of exciton dipoles responding to external electric fields. After removing the backgate voltage the exciton dipoles require time to relax to an equilibrium state, and each dipole's final alignment directions may not be the same as its initial direction. The relationship between exciton dipole directions and PL intensity is discussed later in this study.

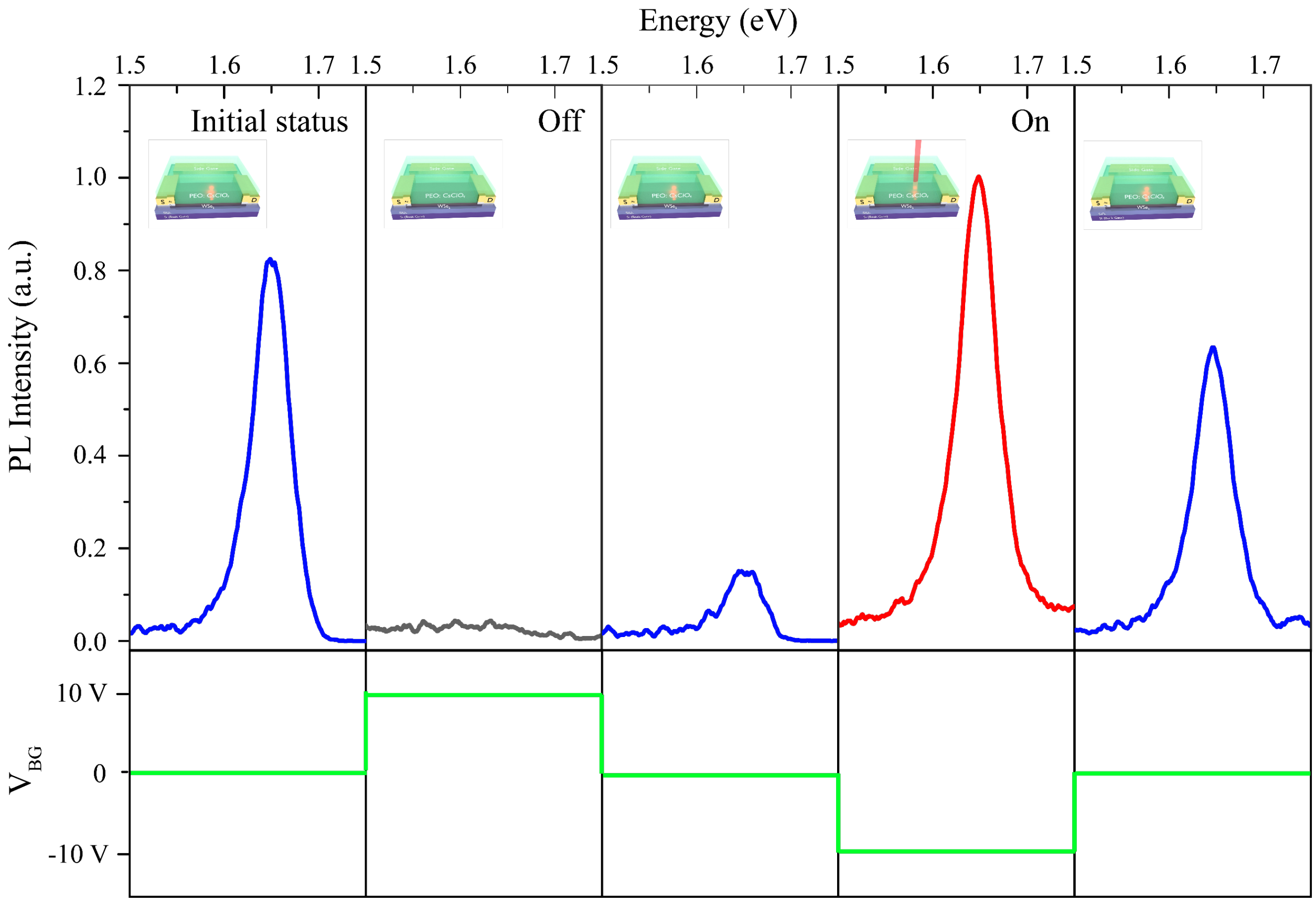

Fig. 2. PL intensity in response to changing $V_{BG}$ showing the switching effect. The top row is the PL spectra of the exfoliated $WSe_2$ transistor in response to a ±10 V back-gate voltage, shown at the bottom.  The side-gate voltage is -2 V.

Figures 3(a) is a schematic of device geometry and exciton dipole arrangements in the absence of electric field (left) and with dual electric fields (right), and Figure 3(b) shows the PL intensity as the function of the gate voltage in both exfoliated and CVD devices. In these measurements, the side-gate was hold at -2 V and the back gate was swept in the range of [-10, 10]. Both samples show the switching effect. The on/off ratio was determined to be 90 and 37 for the exfoliated and CVD grown samples, respectively. When the polarity of the side gate voltage was reversed, the on/off effect can still be observed by tuning the $V_{BG}$ but the on/off ratio decreased, as shown in the supplementary file Figure S4. A control experiment was carried out to investigate the effect of removing the electric field from bottom gate in the CVD-grown sample and only apply the side-gate voltage. As expected, the change of the PL intensity was still observed under the single side-gate conditions, but the on/off ratio was reduced to 12 (Figure S3). This indicates that the PL switching effect is stronger when vertical electric field is applied both on the side gate and back gate.

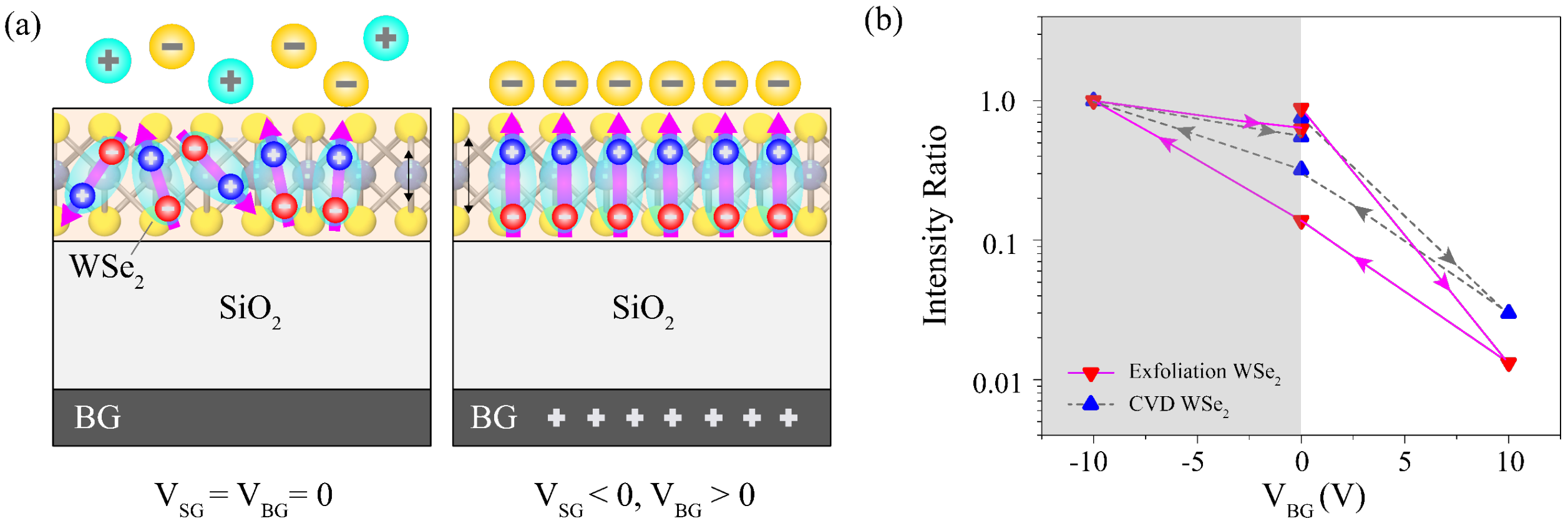

Fig. 3. (a) Schematic of exciton dipole arrangements in the absence of electric field (left) and with dual electric fields (right). Ions are randomly distributed if there is no voltage applied. When a negative $V_{SG}$ is applied, anions accumulate near the channel, while cations (not shown) accumulate near the gate electrode, and bulk part of electrolyte remains charge neutral. (b) PL intensity of the exfoliated and CVD $WSe_2$-based transistors as the function of the back-gate voltage. The side-gate voltage was held at -2 V. The on/off ratios were 90 and 37 respectively. All the data points are normalized by the maximum intensity.

The PL intensity modulation via the gate voltage can be understood as follows. Initially, the excitons in the $WSe_2$ and the ions within the electrolyte are randomly distributed spatially, as illustrated in Figure 3 (a). The applied negative side-gate voltage causes anion accumulation at the interface of the electrolyte and $WSe_2$. In addition to inducing more holes in the $WSe_2$, the anions can also attract holes moving toward the top interface while repelling electrons and moving them in the opposite direction. When a positive back-gate voltage is applied, the separation of the electron and the hole in an exciton is increased, as the positive back-gate pulls the electron towards the bottom and pushes the hole towards the top of the $WSe_2$ layer. The further the electron and the hole are separated spatially, the smaller the overlap of their wavefunction, the lower probability of their recombination. Hence, the PL intensity can be turned off when the two gates possessing the opposite polarity. However, it is turned back on when the polarity of the two gates is tuned the be the same, as there is no longer a force separating the electron and hole pairs within the excitons.

Additionally, when the two gates have the same polarity, they induce the same type of charges in the $WSe_2$ (e.g., holes when $V_{SG} < 0$ V and $V_{BG} < 0$ V), leading to a higher doping density than with only one gate applied (e.g., $V_{SG} < 0$ and $V_{BG} = 0$V). As a result, surface charges can be filled, which reduces the nonradioactive relaxation channels and enhances the PL intensity.[21]

The variation of the PL intensity may also be understood from the change of the population of the excitons which are aligned along the out-of-plane direction. Emission from excitons with dipole moment in the out-of-plane direction is optically forbidden; these excitons are known as "dark" excitons.[24] The dipoles within the $WSe_2$ layer are affected both by the EDL at the electrolyte/$WSe_2$ interface and the back-gate. When an anionic EDL is formed at the $WSe_2$ interface and at the same time a positive back-gate is applied, a unipolar vertical electric field is formed in the monolayer $WSe_2$. It will cause the dipoles in the $WSe_2$ to become vertically aligned and therefore inhibit the emission in the out-of-plane direction. Reversing the polarity of the back-gate voltage changes the direction of the dipoles and thus reduces the populations of the dark excitons in the system. Hence, the PL intensity recovers.

The single and dual-gate measurements on the CVD-grown sample helped test our hypothesis that the PL intensity is strongly affected by the magnitude (Figure S3) and direction of the applied electric field (Figure S4). As mentioned above, the on/off ratio increased from 12 to 37 when an additional gate with same vertical electric field direction was applied. Comparing the dual-gate measurements on CVD-grown and exfoliated samples, the CVD-grown sample has a smaller on/off ratio (37) compared to the exfoliated one (90), which may indicate that the CVD sample contains more impurities than the exfoliated sample. The defects and the impurities in the CVD grown sample can help to bind the carriers from being polarized.

Note that in the voltage range used in this study, the charging time of EDL in the polymer electrolyte is tens to hundreds of seconds;[22] therefore, it would be challenging to conduct a gate-dependent transient PL measurement in this setup. However, our previous study shows that it's possible to reduce the EDL formation time constant to microseconds (or even nanoseconds based on modelling)[22] with significant reduction of device size, which is a potential pathway to better understand the charge carrier dynamics in this type of system. Also, to be noted that because EDL will form at both the electrolyte-$SiO_2$ interface and the electrolyte-$WSe_2$ channel interface, there is some capacitive coupling effect. However, results from finite element modeling using COMSOL Multiphysics (shown in Fig. S5) suggests that capacitive coupling from the nearby oxide doesn't play a significant role in EDL ion density.

We also observed a large Stark shift (blue shift) by controlling the side-gate voltage. This Stark shift was quite different depending on whether the $WSe_2$ layer was exfoliated or CVD grown. The exciton energy of the exfoliated sample does not show an obvious change in response to the electric field, but the CVD-grown sample shows significant modulation, as shown in Figure 4(a). The gate-voltage-dependent blueshift in this case was about 36 meV, which is the largest tunability of this line reported so far.[13,25] The exciton energy in this device shifted upward on both voltage polarities, but by different amounts, which may suggest different surface electric field magnitude. Although different ionic species accumulate at the channel top interface under different polarities (i.e., a cationic EDL forms at $V_{SG} > 0$ V while an anionic EDL forms at $V_{SG} < 0$V), we do not anticipate a change in surface electric field magnitude because our previous experimental results using Hall Effect measurements, transfer measurement,[23,26] and simulation results using finite element modeling[27] suggest that the cationic and anionic EDL have very similar densities. Thus, the shift may be related to the different times required for exciton dipoles to be realigned to a new direction, but more experiments and devices are needed and answering this question is beyond the scope of this study.

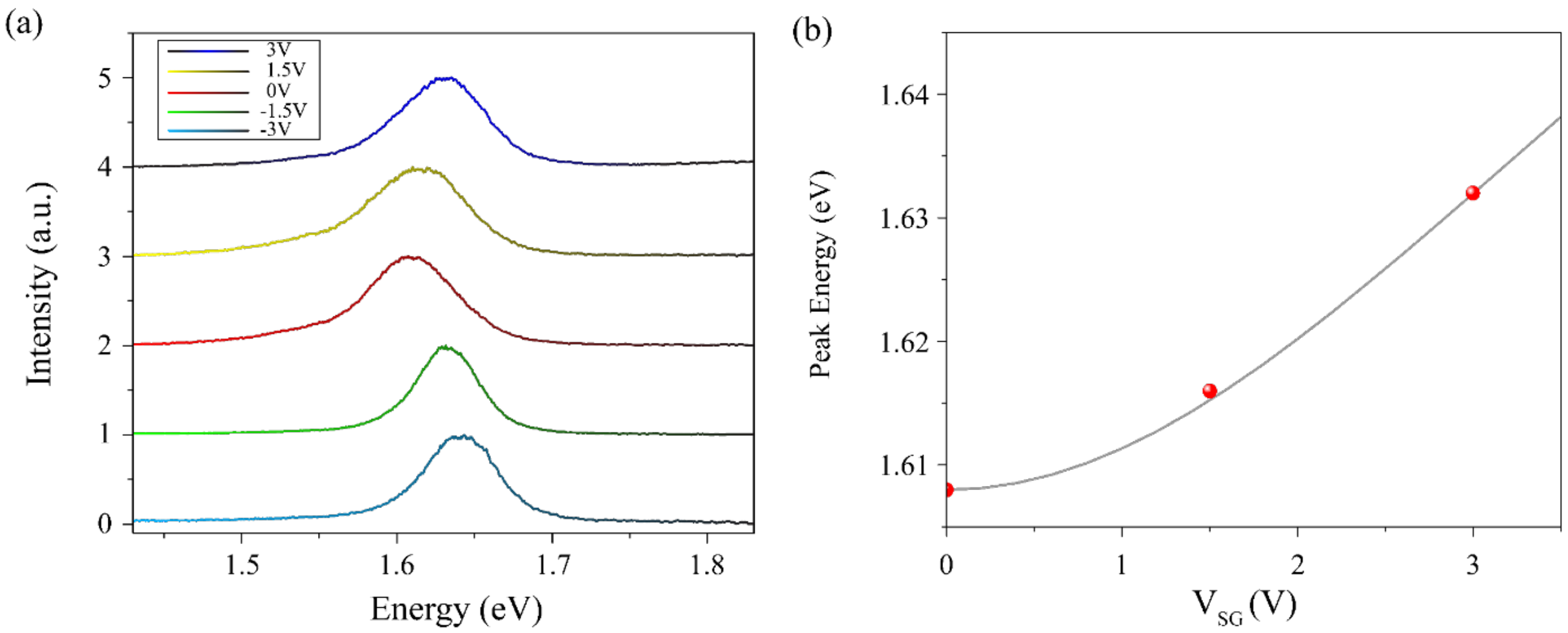

Fig. 4. (a) PL spectra as the function of the side-gate voltage of the CVD-grown $WSe_2$-based transistor. The back-gate was grounded. The exciton energy can be determined by fitting the spectra with the Lorentzian model and $|\Delta E_{max}| = 36$meV is obtained. (b) The peak energy (red dots) of the spectra for the positive polar side-gate overlayed with the calculation (gray line).

The quantum-confined Stark effect in a quantum well can be described by the simple relationship, $\Delta E \propto E^2 d_{QW}^4$,[28,29] where $E$ is electric field strength and the $d_{QW}$ is the thickness of the quantum well. The quantum-confined Stark effect gives a redshift due to the applied electric field. Although the electric field can be approach $10^7$V/cm in our system, due to small thickness of the monolayer $WSe_2$, the electric field should not produce a significant redshift. Similarly, the effect of the wave function distortion in the vertical direction due to the electric field is also negligible, even though it plays a key role in the shift of exciton energy in quantum wells.[30] As discussed above, we have observed a large blueshift of 36 meV in the devices with CVD-grown material. We attribute this blue shift to the nonlinear effect on the in-plane dielectric constant $\epsilon_{||}$ due to the large vertical electric field $E_{\perp}$. In general, the dielectric constant can be expressed as $\epsilon_{||} = \epsilon_0\left(1 + \chi^{(1)} + \chi^{(2)}E_{\perp} + \chi^{(3)}E_{\perp}^2 + \cdots\right)$, where $\chi^{(n)}, n = 1,2,3$ stand for the nth order (off-diagonal) optical susceptibilities. Note that the interaction between the electron and hole has a strong dependence on the in-plane dielectric constant due to the screening effect which in turn modifies the exciton binding energy. Theoretically, this interaction can be modelled by a Keldysh potential[31–36]: $V_K(r) = -\frac{e^2}{8\epsilon_0 r_0}\left[H_0\left(\frac{\kappa r}{r_0}\right) - Y_0\left(\frac{\kappa r}{r_0}\right)\right]$. Here $H_0$ and $Y_0$ are the Struve and Bessel functions of the second kind and the screening length $r_0 = (\epsilon_{\parallel} - 1)d/2$ (with $d = 0.658$ nm being the thickness of the monolayer) largely depends on the dielectric constant. Furthermore, the value of linear susceptibility $\chi^{(1)} \approx 14$ for $WSe_2$ is known in the literature[34,36,37] and the second-order nonlinear susceptibility $\chi^{(2)}$ can be ignored in a uniform medium[38]. Therefore, the lowest-order nonlinear effect is attributed to the third-order optical susceptibility $\chi^{(3)}$, which can be determined by numerically calculating the exciton binding energy and fitting its change with respect to the perpendicular electric field $E_{\perp}$. Our result shows $\chi^{(3)} = 3.50 \times 10^{-19}\ m^2/V^2$, which is the same order of the third-order *diagonal* susceptibility measured in other's experimental work[39] (The details can be referred to the supplementary file). In figure 4(b), we present the measured the PL peak energy along with the theoretically calculated exciton energy as a function of the side-gate voltage. The calculated exciton energy is shifted by a constant in order to match the measured PL peak at zero electric field.

## CONCLUSIONS

Transistors made with $WSe_2$ of two different types show a strong switching of the PL intensity under applied electric field. The mechanism of the intensity variation can be attributed to the change of the background effective doping density and the direction of dipole moment of the excitons as the electric field is varied. This effect may be useful for photonic applications; future work on this system should include detailed studies of the time-dependent behavior to determine the switching speed.

The large modulation of the exciton energy can be explained by the nonlinear effect on the in-plane dielectric constant induced by the vertical electric field. This strong shift may be of benefit when embedding TMD monolayers inside microcavities to make exciton-polaritons, which have great promise for novel photonic devices and Bose-Einstein condensation of exciton-polaritons at room temperature. In general, exciton-polaritons in microcavities are highly sensitive to the exact detuning of the exciton and cavity photon energies, and having direct, electrical control over the detuning provides a very useful design element. While the large binding energy and the strong oscillator strength[40,41] of the excitons in TMD materials are well known, the large tunability of the exciton properties seen here raises many possibilities for all-optical devices at room temperature in the future.[42]

## METHODS

**Device fabrication.** An isolated monolayer $WSe_2$ with a typical size of 5×15 $\mu$m$^2$ was mechanically exfoliated from bulk crystals (purchased from 2D Semiconductors). The whole sample was fabricated by standard dry transferring method to the surface of a p-doped silicon substrate with 90 nm $SiO_2$ coating (purchased from the Graphene Supermarket). Two contacts were written on top of the flake via e-beam lithography, to give a parallel electric field to the surface as the source (S) and drain (D); two side leads were written off the flake for applying the side-gate (SG) voltage. This was followed by an e-beam deposition of Ti/Au 3/100 nm. The metal was grown at a pressure of $10^{-6}$ Torr. We made a similar device upon a CVD-grown sample (material provided by 2DLayer).

**Electrolyte Preparation.** The dual-ion conducting polymer electrolyte (PEO:$CsClO_4$) was prepared similarly to previously published work.[26] PEO (Polymer Standards Service, $M_w$ = 94600 g mol$^{-1}$) and $CsClO_4$ (Sigma-Aldrich, 99.9%) were dissolved in anhydrous acetonitrile (Sigma-Aldrich) to make a 1 wt % solution with an ether oxygen to Cs molar ratio of 20:1 (concentration of salt in PEO is 1177 mol/m$^3$). .

**Optical measurements.** A He-Ne laser with a spot size of 1 μm at normal incidence through an X50 Mitutoyo microscope objective was employed as the pump source. The PL spectra were collected by the same microscope objective and directed towards a charged couple device (CCD) equipped spectroscopy.

**Electronic characterization.** A Keysight B1500A semiconductor parameter analyzer in a Lakeshore cryogenic vacuum probe station (CRX-VF) at the pressure of $2 \times 10^{-6}$ Torr is exploited. For the transfer measurements in Figure 1 the scanning rate is 4 V/s for the back gate and 10 mV/s for the side gate. $V_{DS}$ was 100 mV for both measurements.

## ACKNOLEDGEMENTS

This work was supported by the U.S. Army Research Office under MURI Award No. W911NF-17-1-0312 (Z. S., J. B., and D. S.). Z.S. also acknowledge support from the NSFC Grant No. 12174111, Shanghai Pujiang Program Grant No. 21PJ1403000 and Joint Physics Research Institute Challenge Grant of the NYU-ECNU Institute of Physics at NYU Shanghai. K. X., J. L., and S. K. F.- S. acknowledge funding from NSF-DMR-EPM grant #1607935, and helpful discussions with Brendan Mostek and Shubham Awate. Z.-Y. S. acknowledges support from Program of Shanghai Sailing Program Grant No. 20YF1411600 and NSFC Grant No. 12004115. J. W. acknowledges the support from the National Key R&D Program of China (Grant No. 2018YFA0306303), Shanghai Municipal Science and Technology Major Project.

**Supporting Information Available**

Supplementary includes 1. Optical and electrical characterizations of the as fabricated device; 2. Single and dual gated PL measurements; 3. Optical switching effect for different polarity, for CVD grown $WSe_2$ sample. 4. Theory for nonlinear optical susceptibility and the blue shift of PL energy. 5. Finite element simulation for capacitive coupling effect from electrolyte/oxide interface by COMSOL.

## REFERENCES

(1) Mak, K. F.; Shan, J. Photonics and Optoelectronics of 2D Semiconductor Transition Metal Dichalcogenides. *Nat. Photonics* **2016**, *10* (4), 216–226. https://doi.org/10.1038/nphoton.2015.282.

(2) Xia, F.; Wang, H.; Xiao, D.; Dubey, M.; Ramasubramaniam, A. Two-Dimensional Material Nanophotonics. *Nat. Photonics* **2014**, *8* (12), 899–907. https://doi.org/10.1038/nphoton.2014.271.

(3) Wang, Q. H.; Kalantar-Zadeh, K.; Kis, A.; Coleman, J. N.; Strano, M. S. Electronics and Optoelectronics of Two-Dimensional Transition Metal Dichalcogenides. *Nat. Nanotechnol.* **2012**, *7* (11), 699–712. https://doi.org/10.1038/nnano.2012.193.

(4) Butler, S. Z.; Hollen, S. M.; Cao, L.; Cui, Y.; Gupta, J. A.; Gutiérrez, H. R.; Heinz, T. F.; Hong, S. S.; Huang, J.; Ismach, A. F.; Johnston-Halperin, E.; Kuno, M.; Plashnitsa, V. V.; Robinson, R. D.; Ruoff, R. S.; Salahuddin, S.; Shan, J.; Shi, L.; Spencer, M. G.; Terrones, M.; Windl, W.; Goldberger, J. E. Progress, Challenges, and Opportunities in Two-Dimensional Materials beyond Graphene. *ACS Nano* **2013**, *7* (4), 2898–2926. https://doi.org/10.1021/nn400280c.

(5) Xu, X.; Yao, W.; Xiao, D.; Heinz, T. F. Spin and Pseudospins in Layered Transition Metal Dichalcogenides. *Nat. Phys.* **2014**, *10* (5), 343–350. https://doi.org/10.1038/nphys2942.

(6) Galfsky, T.; Sun, Z.; Considine, C. R.; Chou, C.-T.; Ko, W.-C.; Lee, Y.-H.; Narimanov, E.

E.; Menon, V. M. Broadband Enhancement of Spontaneous Emission in Two-Dimensional Semiconductors Using Photonic Hypercrystals. *Nano Lett.* **2016**, *16* (8). https://doi.org/10.1021/acs.nanolett.6b01558.

(7) Geim, A. K.; Grigorieva, I. V. Van Der Waals Heterostructures. *Nature* **2013**, *499* (7459), 419–425. https://doi.org/10.1038/nature12385.

(8) Liu, X.; Galfsky, T.; Sun, Z.; Xia, F.; Lin, E. C.; Lee, Y. H.; Kéna-Cohen, S.; Menon, V. M. Strong Light-Matter Coupling in Two-Dimensional Atomic Crystals. *Nat. Photonics* **2014**, *9* (1), 30–34. https://doi.org/10.1038/nphoton.2014.304.

(9) Sun, Z.; Gu, J.; Ghazaryan, A.; Shotan, Z.; Considine, C. R.; Dollar, M.; Chakraborty, B.; Liu, X.; Ghaemi, P.; Kéna-Cohen, S.; Menon, V. M. Optical Control of Roomerature Valley Polaritons. *Nat. Photonics* **2017**, *11* (8), 491–496. https://doi.org/10.1038/nphoton.2017.121.

(10) Wu, E.; Xie, Y.; Zhang, J.; Zhang, H.; Hu, X.; Liu, J.; Zhou, C.; Zhang, D. Dynamically Controllable Polarity Modulation of MoTe2 Field-Effect Transistors through Ultraviolet Light and Electrostatic Activation. *Sci. Adv.* **2019**, *5* (5), 1–10. https://doi.org/10.1126/sciadv.aav3430.

(11) Hu, Y.; Huang, Y.; Tan, C.; Zhang, X.; Lu, Q.; Sindoro, M.; Huang, X.; Huang, W.; Wang, L.; Zhang, H. Two-Dimensional Transition Metal Dichalcogenide Nanomaterials for Biosensing Applications. *Mater. Chem. Front.* **2017**, *1* (1), 24–36. https://doi.org/10.1039/c6qm00195e.

(12) Chakraborty, B.; Gu, J.; Sun, Z.; Khatoniar, M.; Bushati, R.; Boehmke, A. L.; Koots, R.; Menon, V. M. Control of Strong Light-Matter Interaction in Monolayer WS2 through Electric Field Gating. *Nano Lett.* **2018**, *18* (10), 6455–6460. https://doi.org/10.1021/acs.nanolett.8b02932.

(13) Sun, Z.; Beaumariage, J.; Xu, K.; Liang, J.; Hou, S.; Forrest, S. R.; Fullerton-Shirey, S. K.; Snoke, D. W. Electric-Field-Induced Optical Hysteresis in Single-Layer WSe2. *Appl. Phys. Lett.* **2019**, *115* (16). https://doi.org/10.1063/1.5123514.

(14) Sun, Z.; Beaumariage, J.; Movva, H. C. P.; Chowdhury, S.; Roy, A.; Banerjee, S. K.; Snoke, D. W. Stress-Induced Bandgap Renormalization in Atomic Crystals. *Solid State Commun.* **2019**, *288* (November 2018), 18–21. https://doi.org/10.1016/j.ssc.2018.11.006.

(15) Johari, P.; Shenoy, V. B. Tuning the Electronic Properties of Semiconducting Transition

Metal Dichalcogenides by Applying Mechanical Strains. *ACS Nano* **2012**, *6* (6), 5449–5456. https://doi.org/10.1021/nn301320r.

(16) Feng, J.; Qian, X.; Huang, C. W.; Li, J. Strain-Engineered Artificial Atom as a Broad-Spectrum Solar Energy Funnel. *Nat. Photonics* **2012**, *6* (12), 866–872. https://doi.org/10.1038/nphoton.2012.285.

(17) Ye, J. T.; Zhang, Y. J.; Akashi, R.; Bahramy, M. S.; Arita, R.; Iwasa, Y. Superconducting Dome in a Gate-Tuned Band Insulator. *Science (80-. ).* **2012**, *338* (6111), 1193–1196. https://doi.org/10.1126/science.1228006.

(18) Bisri, S. Z.; Shimizu, S.; Nakano, M.; Iwasa, Y. Endeavor of Iontronics: From Fundamentals to Applications of Ion-Controlled Electronics. *Adv. Mater.* **2017**, *29* (25), 1–48. https://doi.org/10.1002/adma.201607054.

(19) Xu, K.; Fullerton-Shirey, S. K. Electric-Double-Layer-Gated Transistors Based on Two-Dimensional Crystals: Recent Approaches and Advances. *J. Phys. Mater.* **2020**, *3* (3), 032001. https://doi.org/10.1088/2515-7639/ab8270.

(20) Li, Z.; Chang, S. W.; Chen, C. C.; Cronin, S. B. Enhanced Photocurrent and Photoluminescence Spectra in MoS2 under Ionic Liquid Gating. *Nano Res.* **2014**, *7* (7), 973–980. https://doi.org/10.1007/s12274-014-0459-2.

(21) Rhodes, D.; Chae, S. H.; Ribeiro-Palau, R.; Hone, J. Disorder in van Der Waals Heterostructures of 2D Materials. *Nat. Mater.* **2019**, *18* (6), 541–549. https://doi.org/10.1038/s41563-019-0366-8.

(22) Xu, K.; Islam, M. M.; Guzman, D.; Seabaugh, A. C.; Strachan, A.; Fullerton-Shirey, S. K. Pulse Dynamics of Electric Double Layer Formation on All-Solid-State Graphene Field-Effect Transistors. *ACS Appl. Mater. Interfaces* **2018**, *10* (49), 43166–43176. https://doi.org/10.1021/acsami.8b13649.

(23) Li, H. M.; Xu, K.; Bourdon, B.; Lu, H.; Lin, Y. C.; Robinson, J. A.; Seabaugh, A. C.; Fullerton-Shirey, S. K. Electric Double Layer Dynamics in Poly(Ethylene Oxide) LiClO4 on Graphene Transistors. *J. Phys. Chem. C* **2017**, *121* (31), 16996–17004. https://doi.org/10.1021/acs.jpcc.7b04788.

(24) Zhou, Y.; Scuri, G.; Wild, D. S.; High, A. A.; Dibos, A.; Jauregui, L. A.; Shu, C.; De Greve, K.; Pistunova, K.; Joe, A. Y.; Taniguchi, T.; Watanabe, K.; Kim, P.; Lukin, M. D.; Park, H. Probing Dark Excitons in Atomically Thin Semiconductors via Near-Field

Coupling to Surface Plasmon Polaritons. *Nat. Nanotechnol.* **2017**, *12* (9), 856–860. https://doi.org/10.1038/nnano.2017.106.

(25) Matsuki, K.; Pu, J.; Kozawa, D.; Matsuda, K.; Li, L. J.; Takenobu, T. Effects of Electrolyte Gating on Photoluminescence Spectra of Large-Area WSe2 Monolayer Films. *Jpn. J. Appl. Phys.* **2016**, *55* (6). https://doi.org/10.7567/JJAP.55.06GB02.

(26) Xu, K.; Liang, J.; Woeppel, A.; Bostian, M. E.; Ding, H.; Chao, Z.; McKone, J. R.; Beckman, E. J.; Fullerton-Shirey, S. K. Electric Double-Layer Gating of Two-Dimensional Field-Effect Transistors Using a Single-Ion Conductor. *ACS Appl. Mater. Interfaces* **2019**, *11* (39), 35879–35887. https://doi.org/10.1021/acsami.9b11526.

(27) Woeppel, A.; Xu, K.; Kozhakhmetov, A.; Awate, S.; Robinson, J. A.; Fullerton-Shirey, S. K. Single- versus Dual-Ion Conductors for Electric Double Layer Gating: Finite Element Modeling and Hall-Effect Measurements. *ACS Appl. Mater. Interfaces* **2020**, *12* (36), 40850–40858. https://doi.org/10.1021/acsami.0c08653.

(28) Bastard, G.; Mendez, E. E.; Chang, L. L.; Esaki, L. Variational Calculations on a Quantum Well in an Electric Field. *Phys. Rev. B* **1983**, *28* (6), 3241–3245. https://doi.org/10.1103/PhysRevB.28.3241.

(29) Singh, J. Electronic and Optoelectronic Properties of Semiconductor Structures [Book Review]. *Cambridge Univ. Press* **2003**. https://doi.org/doi:10.1017/CBO9780511805745.

(30) Szymanska, M. H.; Littlewood, P. B. Excitonic Binding in Coupled Quantum Wells. *Phys. Rev. B - Condens. Matter Mater. Phys.* **2003**, *67* (19), 3–6. https://doi.org/10.1103/PhysRevB.67.193305.

(31) Stier, A. V.; Wilson, N. P.; Velizhanin, K. A.; Kono, J.; Xu, X.; Crooker, S. A. Magnetooptics of Exciton Rydberg States in a Monolayer Semiconductor. *Phys. Rev. Lett.* **2018**, *120* (5), 1–6. https://doi.org/10.1103/PhysRevLett.120.057405.

(32) Keldysh, L. V. Coulomb Interaction in Thin Semiconductor and Semimetal Films. JETP Letters 1979, pp 658–661.

(33) Cudazzo, P.; Tokatly, I. V.; Rubio, A. Dielectric Screening in Two-Dimensional Insulators: Implications for Excitonic and Impurity States in Graphane. *Phys. Rev. B - Condens. Matter Mater. Phys.* **2011**, *84* (8), 1–7. https://doi.org/10.1103/PhysRevB.84.085406.

(34) Berkelbach, T. C.; Hybertsen, M. S.; Reichman, D. R. Theory of Neutral and Charged

Excitons in Monolayer Transition Metal Dichalcogenides. *Phys. Rev. B - Condens. Matter Mater. Phys.* **2013**, *88* (4), 1–6. https://doi.org/10.1103/PhysRevB.88.045318.

(35) Wu, F.; Qu, F.; Macdonald, A. H. Exciton Band Structure of Monolayer MoS2. *Phys. Rev. B - Condens. Matter Mater. Phys.* **2015**, *91* (7), 1–8. https://doi.org/10.1103/PhysRevB.91.075310.

(36) Kylänpää, I.; Komsa, H. P. Binding Energies of Exciton Complexes in Transition Metal Dichalcogenide Monolayers and Effect of Dielectric Environment. *Phys. Rev. B - Condens. Matter Mater. Phys.* **2015**, *92* (20), 1–6. https://doi.org/10.1103/PhysRevB.92.205418.

(37) Stier, A. V.; Wilson, N. P.; Clark, G.; Xu, X.; Crooker, S. A. Probing the Influence of Dielectric Environment on Excitons in Monolayer WSe2: Insight from High Magnetic Fields. *Nano Lett.* **2016**, *16* (11), 7054–7060. https://doi.org/10.1021/acs.nanolett.6b03276.

(38) Boyd, R. W. *Nolinear Optics*; 1967.

(39) Autere, A.; Jussila, H.; Marini, A.; Saavedra, J. R. M.; Dai, Y.; Säynätjoki, A.; Karvonen, L.; Yang, H.; Amirsolaimani, B.; Norwood, R. A.; Peyghambarian, N.; Lipsanen, H.; Kieu, K.; De Abajo, F. J. G.; Sun, Z. Optical Harmonic Generation in Monolayer Group-VI Transition Metal Dichalcogenides. *Phys. Rev. B* **2018**, *98* (11), 1–7. https://doi.org/10.1103/PhysRevB.98.115426.

(40) Qiu, D. Y.; Da Jornada, F. H.; Louie, S. G. Optical Spectrum of MoS2: Many-Body Effects and Diversity of Exciton States. *Phys. Rev. Lett.* **2013**, *111* (21), 1–5. https://doi.org/10.1103/PhysRevLett.111.216805.

(41) Chernikov, A.; Berkelbach, T. C.; Hill, H. M.; Rigosi, A.; Li, Y.; Aslan, O. B.; Reichman, D. R.; Hybertsen, M. S.; Heinz, T. F. Exciton Binding Energy and Nonhydrogenic Rydberg Series in Monolayer WS2. *Phys. Rev. Lett.* **2014**, *113* (7), 1–5. https://doi.org/10.1103/PhysRevLett.113.076802.

(42) Sun, Z.; Snoke, D. W. Optical Switching with Organics. *Nat. Photonics* **2019**, *13* (June), 370–371.

# Supplementary information for Photoluminescence switching effect in a two-dimensional atomic crystal

Zheng Sun[1,2*‡], Ke Xu[3,4,5‡], Chang Liu[1,6‡] Jonathan Beaumariage[2], Jierui Liang[2], Susan K Fullerton-Shirey[3], Zhe-Yu Shi[1], Jian Wu[1,7,8], and David Snoke[2]

[1]State Key Laboratory of Precision Spectroscopy, East China Normal University, Shanghai, 200241, China

[2]Department of Physics and Astronomy, University of Pittsburgh, Pittsburgh, PA 15260, USA

[3]Department of Chemical and Petroleum Engineering, University of Pittsburgh, Pittsburgh, PA 15260, USA

[4]School of Physics and Astronomy, Rochester Institute of Technology, 14623, USA

[5]Microsystems Engineering, Rochester Institute of Technology, 14623, USA

[6]Institute for Advanced Study, Tsinghua University, Beijing 100084, China

[7]Collaborative Innovation Center of Extreme Optics, Shanxi University, Taiyuan, Shanxi 030006, China

[8]CAS Center for Excellence in Ultra-intense Laser Science, Shanghai 201800, China

[*] Email: zsun@lps.ecnu.edu.cn.
[‡] Contributed equally to this work.

## I. Optical microscope image of CVD sample:

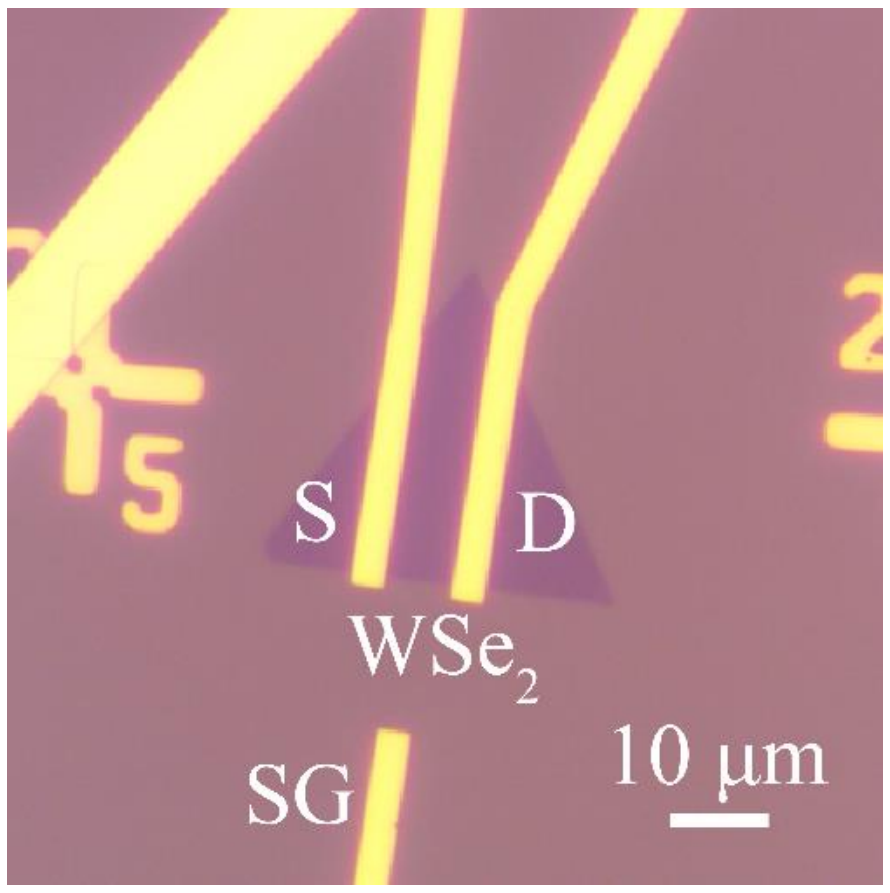

Fig. S1. Optical microscope image of CVD grown transistor with a side-gate. The tri-angular domain is the monolayer regime.

## II. Transfer characteristics of CVD sample prior to electrolyte deposition.

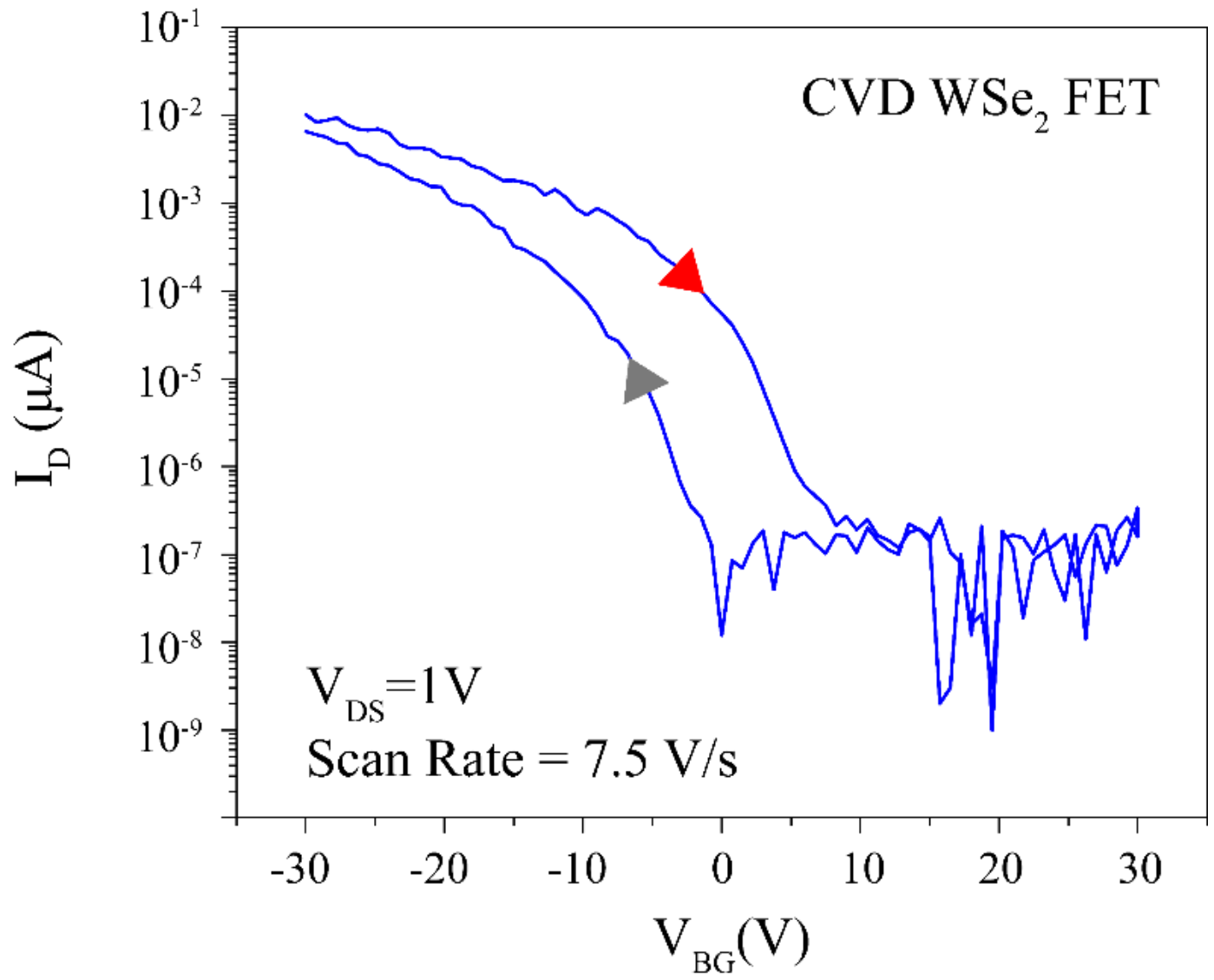

Fig. S2. Back-gate transfer characteristics of CVD $WSe_2$ FET prior to electrolyte deposition. Only a p-branch is observed within the measurement window, indicating p-type doping of the as-grown material.

### III. PL intensity modulation using single vs dual gates on CVD sample

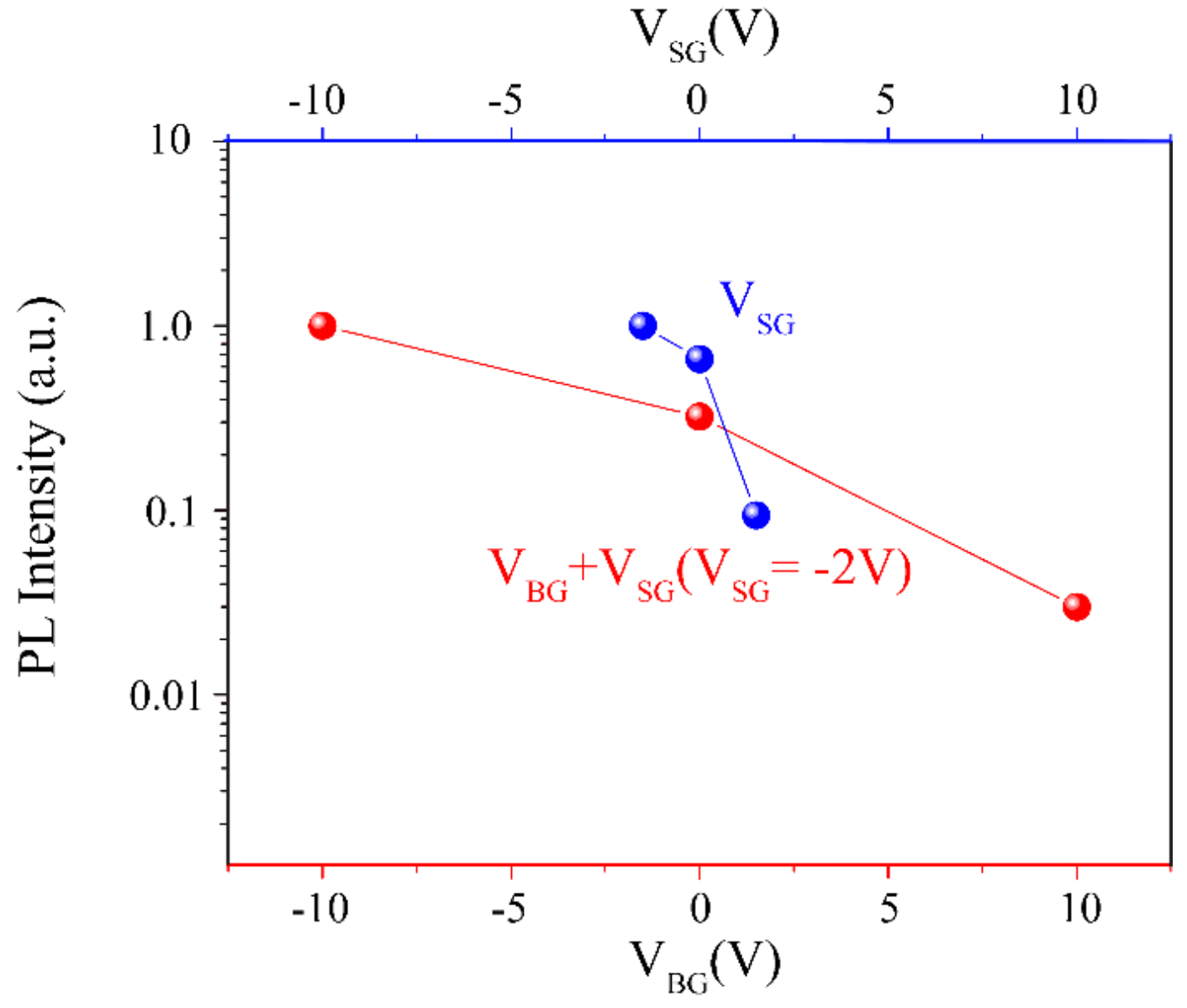

Fig. S3. PL intensity modulation with only side gate (blue) and dual gates (red). An on/off ratio of 12 is observed with only the side-gate, and the ratio is of 37 for the dual gate, in agreeing with our discussion of the mechanisms of PL intensity modulation.

Fig. S3 gives the results of a control experiment which was carried out to decouple the effect of the top and bottom electric field in the CVD-grown sample, by applying only the side-gate voltage and the dual gates. As expected, the change of the PL intensity can still be observed under the single side-gate conditions, but the on/off ratio is reduced to 12. These results suggest that the PL switching effect is stronger when both gates are applied with vertical electric fields aligned in the same direction.

## IV. Optical switching for different polarity.

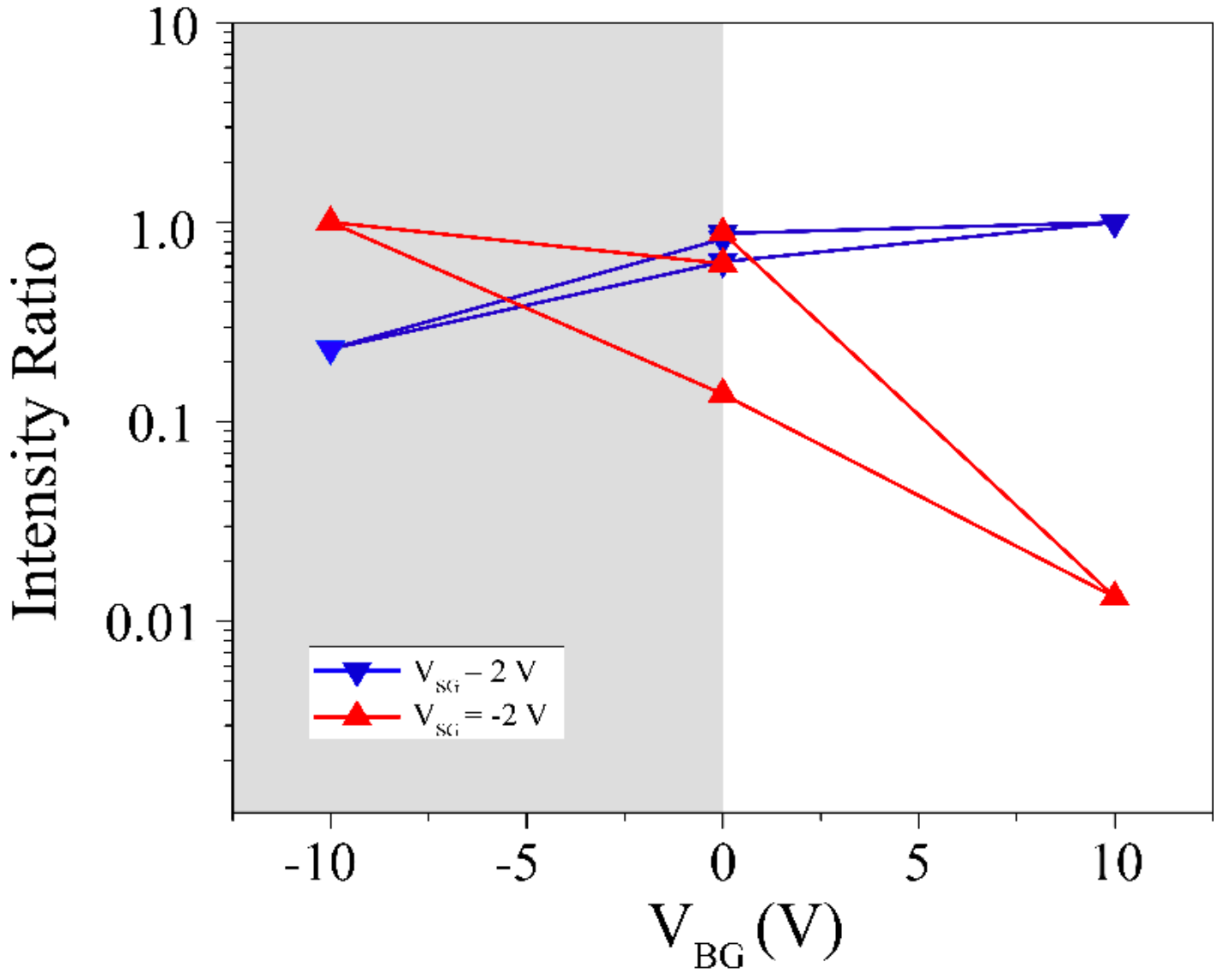

Fig. S4 PL intensity of the CVD WSe2-based transistors as the function of the back-gate voltage with initially set the side-gate voltage of -2 V (in red) and 2 V (in blue). The on/off ratios were 37 and 7 respectively. All the data points are normalized by the maximum intensity.

As discussed in the manuscript, the device turns off when the EDL ($V_{SG}$) induced electric field is in the same direction as $V_{BG}$ induced field. This is still true when side gate voltage is changed from -2V to +2V, and the device turns off when $V_{BG}$ is negative and have the same direction of $V_{SG}$ field. However, the on/off ratio when $V_{SG}$ is +2V is about 7, lower than the ratio of 37 when $V_{SG}$ is -2V. We don't think it's majorly the impact of ionic species because our previous experimental results using Hall Effect measurement and transfer measurement,[1,2] and simulation results using finite element modeling[3] both suggest that the cationic and anionic EDL have very similar density. More devices and experiments would be needed to have a conclusive understanding. One possible explanation for this difference in PL on/off ratio might be the intrinsic p-type doping of the CVD grown WSe2 sample, which can be seen from the transfer measurement in Fig. S2. At the positive voltage the current, therefore charge carrier density, is lower than at the negative voltage, which may indicate we need different voltages to align exciton dipoles to turn on/off PL. It is an interesting effect that we can investigate in a future study.

## V. Nonlinear optical susceptibility and the blue shift of PL energy

The screened Coulomb interaction in a monolayer material is given by the Keldysh potential[4–9],

$$V_K(r) = -\frac{e^2}{8\epsilon_0 r_0}\left[H_0\left(\frac{\kappa\, r}{r_0}\right) - Y_0\left(\frac{\kappa\, r}{r_0}\right)\right],$$

where the screening length $r_0 = (\epsilon_{\parallel} - 1)d/2$ depends on the layer separation d and in-plane dielectric constant $\epsilon_{||}$, and the average dielectric constant $\kappa = (\epsilon_{top} + \epsilon_{bottom})/2$ (we take $\kappa = 4.5$ as the same value taken in the literature[4–9]).

Under a strong vertical electric field, the dielectric constant can be expressed as $\epsilon_{||} = \epsilon_0\left(1 + \chi^{(1)} + \chi^{(2)}E_{\perp} + \chi^{(3)}E_{\perp}^2 + \cdots\right)$, where $\chi^{(n)}, n = 1,2,3$ stand for the nth order (off-diagonal) optical susceptibilities. In a uniform medium, the second order nonlinear susceptibility $\chi^{(2)}$ is ignored[10] and we only consider the third order nonlinear susceptibility $\chi^{(3)}$. The screening length $r_0$ then has a quadratic dependence on the vertical electric field because of $\chi^{(3)}$, which causes the change of exciton binding energy and eventually leads to the blue shift of the PL signal.

We numerically calculated the binding energy for different screening length $r_0$ according via the two-body Schrödinger equation,

$$\frac{\hbar^2\nabla^2}{2m_r}\phi(\mathbf{r}) + V_K(r) = E\phi(\mathbf{r})$$

with $m_r = 0.2m_e$ being the reduced effective mass of electron and hole. The binding energy is then fitted with the experimental measured PL blueshift and gives the off-diagonal optical susceptibility $\chi^{(3)} = 3.50 \times 10^{-19}\ m^2/V^2$.

## VI. Capacitive coupling effect from electrolyte/oxide interface

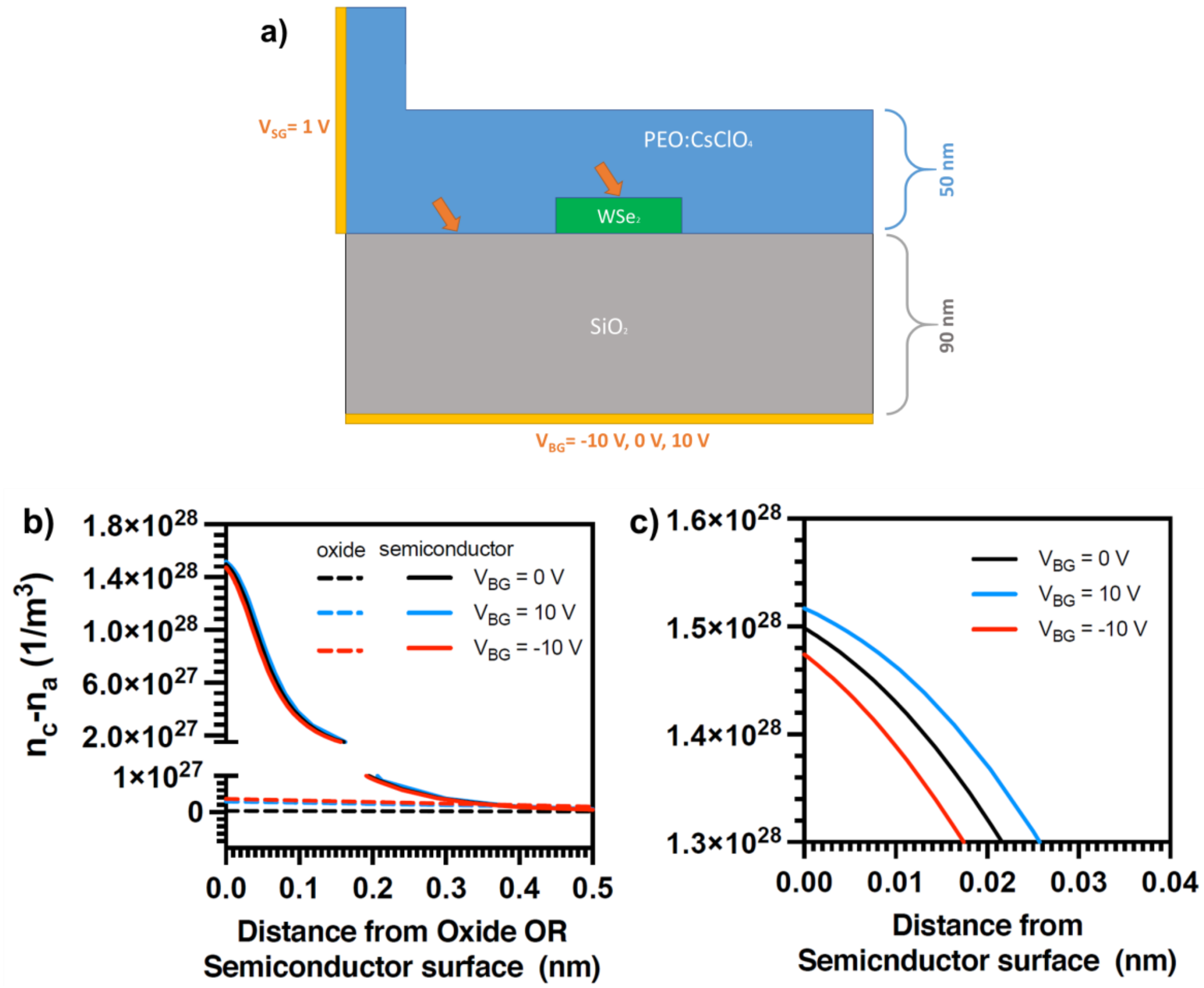

Fig. S5. a) Side gate device geometry setup in COMSOL. The arrows indicate locations where ion density are plotted in b) and c). b) Ion density as a function of distance from both the $WSe_2$ (solid lines) and oxide surfaces (dashed lines) into the electrolyte at $V_{BG}$ = 0, -10, +10 V. c) Zoomed in view of EDL ion density at the electrolyte/$WSe_2$ interface.

Because EDL will form at both the electrolyte-$SiO_2$ interface and the electrolyte-$WSe_2$ channel interface, there is some capacitive coupling effect. However, to quantify whether or not this coupling is significant, we ran additional finite element modeling using COMSOL Multiphysics. We set the side gate equal to 1 V and the backgate to -10, 0 or +10 V. A schematic of the side-gated device geometry and ion density as a function of distance from both the $WSe_2$ and oxide surfaces into the electrolyte are provided in Fig. S5. The ion density in the EDL at the surface of the semiconductor is two orders of magnitude larger than that at the electrolyte/oxide interface. With the $V_{BG}$ set to +10 V and the $V_{SG}$ set to +1 V (i.e., the fields reinforce each other and would give the highest possible capacitive coupling), the EDL concentration on top of the $WSe_2$ is 1.3% larger than when the $V_{BG}$ is set to zero. A 1.3% modulation on the $WSe_2$ with the backgate shows that the $V_{BG}$ and the channel EDL are mostly independent of each other, and that capacitive coupling from the nearby oxide doesn't play a significant role.

**Reference:**

(1) Li, H. M.; Xu, K.; Bourdon, B.; Lu, H.; Lin, Y. C.; Robinson, J. A.; Seabaugh, A. C.; Fullerton-Shirey, S. K. Electric Double Layer Dynamics in Poly(Ethylene Oxide) LiClO4 on Graphene Transistors. *J. Phys. Chem. C* **2017**, *121* (31), 16996–17004. https://doi.org/10.1021/acs.jpcc.7b04788.

(2) Xu, K.; Liang, J.; Woeppel, A.; Bostian, M. E.; Ding, H.; Chao, Z.; McKone, J. R.; Beckman, E. J.; Fullerton-Shirey, S. K. Electric Double-Layer Gating of Two-Dimensional Field-Effect Transistors Using a Single-Ion Conductor. *ACS Appl. Mater. Interfaces* **2019**, *11* (39), 35879–35887. https://doi.org/10.1021/acsami.9b11526.

(3) Woeppel, A.; Xu, K.; Kozhakhmetov, A.; Awate, S.; Robinson, J. A.; Fullerton-Shirey, S. K. Single- versus Dual-Ion Conductors for Electric Double Layer Gating: Finite Element Modeling and Hall-Effect Measurements. *ACS Appl. Mater. Interfaces* **2020**, *12* (36), 40850–40858. https://doi.org/10.1021/acsami.0c08653.

(4) Stier, A. V.; Wilson, N. P.; Velizhanin, K. A.; Kono, J.; Xu, X.; Crooker, S. A. Magnetooptics of Exciton Rydberg States in a Monolayer Semiconductor. *Phys. Rev. Lett.* **2018**, *120* (5), 1–6. https://doi.org/10.1103/PhysRevLett.120.057405.

(5) Keldysh, L. V. Coulomb Interaction in Thin Semiconductor and Semimetal Films. JETP Letters 1979, pp 658–661.

(6) Cudazzo, P.; Tokatly, I. V.; Rubio, A. Dielectric Screening in Two-Dimensional Insulators: Implications for Excitonic and Impurity States in Graphane. *Phys. Rev. B - Condens. Matter Mater. Phys.* **2011**, *84* (8), 1–7. https://doi.org/10.1103/PhysRevB.84.085406.

(7) Berkelbach, T. C.; Hybertsen, M. S.; Reichman, D. R. Theory of Neutral and Charged Excitons in Monolayer Transition Metal Dichalcogenides. *Phys. Rev. B - Condens. Matter*

*Mater. Phys.* **2013**, *88* (4), 1–6. https://doi.org/10.1103/PhysRevB.88.045318.

(8) Wu, F.; Qu, F.; Macdonald, A. H. Exciton Band Structure of Monolayer MoS2. *Phys. Rev. B - Condens. Matter Mater. Phys.* **2015**, *91* (7), 1–8. https://doi.org/10.1103/PhysRevB.91.075310.

(9) Kylänpää, I.; Komsa, H. P. Binding Energies of Exciton Complexes in Transition Metal Dichalcogenide Monolayers and Effect of Dielectric Environment. *Phys. Rev. B - Condens. Matter Mater. Phys.* **2015**, *92* (20), 1–6. https://doi.org/10.1103/PhysRevB.92.205418.

(10) Boyd, R. W. *Nolinear Optics*; 1967.